\documentclass[12pt,reqno]{article}
\usepackage{amsmath,amssymb,amsfonts,mathtools} 
\usepackage{amssymb}  	 
\usepackage{makeidx}  	 
\usepackage{graphicx} 		 
\usepackage{epsfig}
\usepackage{appendix}		
\usepackage{empheq}
\usepackage{setspace} 	
\usepackage{enumitem}
\usepackage[numbers,sort&compress]{natbib}
\usepackage{pstricks}
\usepackage{soul}  
\usepackage[dvips]{curves}  
\usepackage{xcolor}
\usepackage{listings}
\usepackage{amsbsy}

\usepackage{multicol}
\usepackage{booktabs}
\usepackage{verse}
\usepackage{pstricks, pst-text}                  
\usepackage{color}
\usepackage{fancyhdr}
\usepackage[T1]{fontenc}						
\usepackage{newpxtext,newpxmath}   	
\usepackage{titlesec}
\titleformat{\section}{\normalfont\fontsize{12}{13}\bfseries}{\thesection}{6pt}{\setstretch{0.1}}
  \titleformat{\subsection}
  {\normalfont\fontsize{11}{12}\bfseries}{\thesubsection}{1em}{}
  \titleformat{\subsubsection} 
  {\normalfont\fontsize{11}{11}\bfseries}{\thesubsubsection}{1em}{}

\newcommand\myfigure[5]{%
  \ifdim#2>.8\linewidth
    {%
      \centering
      \includegraphics[width=#3]{#4}%
      \captionof{figure}{#5}%
    }%
  \else
  \begin{wrapfigure}{#1}{#2}
    \includegraphics[width=#3]{#4}
    \caption{#5}
  \end{wrapfigure}
  \fi
}

\usepackage{graphicx}
\usepackage{wrapfig}
\usepackage{caption}
\newcommand{\linkcolor}{blue}
\usepackage[colorlinks=true,linkcolor=\linkcolor,urlcolor=\linkcolor]{hyperref} 
\usepackage{academicons}
\definecolor{orcidlogocol}{HTML}{A6CE39}

\begin{document}
\noindent
\textbf{\Large Confusion concerning the extrapolated endpoint. When will it ever end?}\\[2em]
K. Razi Naqvi (\url{razi.naqvi@ntunu.no}; \url{https://orcid.org/0000-0002-9776-1249})\\
\textit{Department of Physics, Norwegian University of Science and Technology (NTNU), 7094 Trondheim, Norway}\\
\vspace*{2ex}

\leftskip=20truept
\rightskip=20truept
\setstretch{0.9}
{\small
\noindent
In a paper on ``the Brownian motion analog of the well-known Milne problem in radiative transfer theory'' [\textit{J Stat Phys} 25 (1981) 569--82], Burschka and Titulaer reported: ``The value we find for this `Milne extrapolation length' is, in the appropriate dimensionless units, approximately twice the value found in the radiative transfer problem.'' A study by Ziff [\textit{J Stat Phys} 65 (1991) 1217--33], concerned with the absorption of particles executing a Rayleigh flight (randomly directed displacements of equal length $l$) by a black sphere of radius $R$, led to a value for the extrapolation length $\gamma$ about half as small as the benchmark result (for $R\gg l)$. The first discrepancy is shown to result from the disparity of the two length scales; the second, from the zero variance of the jump lengths. Ziff's finding that $\gamma$ is independent of $R$ for $0<l\leq 2R$, cannot be reconciled with studies based on the Lorentz-Boltzmann equation and the Klein-Kramers equation.} \\[1em]



\setstretch{1.04}
\leftskip=0truept
\rightskip=0truept
\vspace{4em}

\begin{multicols}{2}

\section{Introduction}
\label{sec:Intro}

The purpose of this article is to ask which if either of the discrepancies mentioned in the abstract is  genuine, and to probe into their implications for the long-established (but ill-advised) practice of equating the rate constant of diffusion-mediated bimolecular reactions in liquids to the flux found by solving the diffusion equation (DE) subject to the Smoluchowski boundary condition (SBC), and multiplying the calculated flux--if found to be smaller than the experimentally determined rate constant--by a factor smaller than unity \cite{Smoluchowski1916PhysZeit,Smoluchowski1916ZPC,Smoluchowski1917KZ}.\par

\medskip
The background can be best laid out by quoting the introductory lines from a 1981 article \cite{Burschka+Titulaer1981JSPb} of Burschka and Titulaer (B\&T):\par
\smallskip
\leftskip=20truept
\setstretch{0.9}
{\small\noindent
The flow of a reactant in a diffusion-controlled reaction can often be described in terms of Brownian motion of a particle in the presence of absorbing or partially absorbing boundaries. The simplest description is obtained through a diffusion or Smoluchowski equation for the probability density of the particle position with either absorbing or ``radiative'' boundary conditions. In the former case the density is put equal to zero at the boundary, while in the latter the outward normal flux is proportional to the density with a phenomenological proportionality constant. This traditional theory has often been criticized; in particular there seems to be no clear way of relating the proportionality constant in the radiative boundary conditions to a microscopic picture of the reaction kinetics.}\par
{\small The reasons for the inadequacy of the Smoluchowski equation can be exhibited by inspecting its derivation from a more detailed description of Brownian motion due to Klein and Kramers, in terms of the probability density for the velocity and position of the Brownian particle. The Smoluchowski equation can be recovered from this description via a procedure of the Chapman-Enskog type. This derivation breaks down, however, near a wall or at places where the potential varies rapidly; there a so-called kinetic boundary layer occurs. This breakdown is caused by the large deviations from the Maxwellian velocity distribution that must occur, e.g., near an absorbing boundary, whereas for validity of the Smoluchowski equation approximate local equilibrium is required. [Bibliographic indicators have been suppressed here.]}\par
\medskip
\leftskip=0pt
\setstretch{1.04}
\leftskip=0truept
\smallskip 
The two paragraphs quoted above provide an accurate account of the (then) prevalent paradigm. B\&T could not have foreseen that new evidence, sufficient to overturn the paradigm, was in the offing. The evidence \cite{KRN+KJM+SW1982JCP,KRN+KJM+SW1982PRL,KRN+SW+KJM1982JPC,KRN+SW+KJM1983JCP}, which became available soon after the publication of ref~\cite{Burschka+Titulaer1981JSPb}, failed to deflate the paradigm, and caused at most a few minute punctures, which appeared to have been repaired by the dust which accumulates on printed matter and memory cells, if not blown away every now and again by enquiring spirits. Some comments on the passage are therefore in order, and should be construed not as shooting the messenger but as an attempt to shoot down a paradigm that has proved to be particularly recalcitrant.

Although an additional and shorter excerpt from ref~\cite{Burschka+Titulaer1981JSPb} is needed, a statement of the Milne problem (tailored for this note), a few nomenclatory notes, and a recapitulation of the results most pertinent for the present discussion must precede the last excerpt.\par

\section{Preliminary material}
\subsection{Statement of the Milne problem and terminology}
\label{subsec:Milne}
A homogeneous, semi-infinite, non-absorbing medium occupies the half-space $x>0$, and sustains a constant current of test particles in the negative $x$ direction. The medium (or the \textit{host}) itself contains no sources or sinks, and the plane boundary at $x=0$ acts as a black wall that absorbs all particles incident on it. The test particles obey either the one-speed Lorentz-Boltzmann equation (LBE) of neutron transport or radiative transfer, in which case one is faced with the Milne problem, or the Klein-Kramers equation (KKE), which leads one to the Brownian analog of the Milne problem. The \textit{problem} is to determine $n(x)$, the density of the particles inside the medium ($x>0$).

As two versions of the KKE have been brought to bear on the Milne problem, the label \textit{Klein-Kramers equation} needs some elaboration. When Fermi analyzed Milne's problem, in the context of neutron diffusion \cite[pp.~980--1016]{Fermi1962CollPap1}, he decided to simplify the analysis ``by considering the fictitious case in which the neutrons move in one dimension along a line, instead of being free in space.'' The choice, he explained, ``has the the advantage that we can easily obtain approximate expressions valid for the actual three-dimensional case.'' The equation used by B\&T applies to true one-dimensional Brownian motion. The Trondheim group (myself and two colleagues) have investigated this version \cite{KRN+KJM+SWPhysRevA1989} as well as the complete analog of the Milne problem, in which particles move in three-dimensional space, but plane symmetry prevails and the cocentration depends on only one cartesian coordinate \cite{SW+KJM+KRN1983PhysRevA}. Menon, Kumar and Sahni \cite{Menon+Kumar+Sahni1986Physica} pointed out that, as regards particle density, the two versions stand on an equal footing, but their difference does manifest itself when other physical moments are computed. 

A ``gas-phase Brownian analog'' of the Milne's problem, examined by Lindenfeld and Shizgal (L\&S) \cite{LindShiz1983MilneMass}, should also be mentioned at this point. They analyzed, within the framework of the Boltzmann equation, a variant of the Milne problem for a gaseous mixture consisting of rigid elastic spheres, in which the number density $\overline{n}_1$ of the host particles (at a given temperature $T_1$) far exceeds $\overline{n}$, the number density of the test particles. A crucial parameter in this context is the mass ratio $\gamma{\hspace{-3pt}\smash[t]{\strut}}_{m}=m_1/m$, where $m_1$ is the mass of the host particles, and $m$ that of the test particles. In the terminology of Andersen and Shuler \cite{Andersen+Shuler1964}, the system becomes a ``Lorentz gas'' when $\gamma_m\gg 1$,  and a "Rayleigh gas'' when $\gamma{\hspace{-3pt}\smash[t]{\strut}}_{m}\ll 1$. In the limit $\gamma{\hspace{-3pt}\smash[t]{\strut}}_{m}\to \infty$, the problem considered by L\&S coincides with that formulated by the eponym, Milne, and their results agree well with the exact calculations; commenting on the opposite extreme $\gamma{\hspace{-3pt}\smash[t]{\strut}}_{m}\to 0$, L\&S remarked: ``It might be of interest to compare this result with a calculation employing the Fokker-Planck [KK] operator which coincides with the Boltzmann collision operator in this limit.'' The results of an approximate analytical treatment of the plane-symmetric form of the KKE, carried out by the Trondheim group \cite{SW+KJM+KRN1983PhysRevA},  accorded well with the calculations of L\&S.

Let us recall the nomenclature common in neutron transport literature \cite[p.~73]{Davison1957Neutron}. The symbol $x_0$ will denote the \textit{extrapolated endpoint}, the point beyond the boundary ($x=0$) at which the extrapolated part of the asymptotic density $n_{\rm as}(x)$ vanishes:
\begin{equation}\label{eq:def-x0}
n_{\rm as}(-x_0)=0.
\end{equation}
The \textit{linear extrapolation length} $l_{\rm ex}$ is defined through the relation:
\begin{equation}\label{eq:l-ext}
l_{\rm ex}=\frac{n_{\rm as}(0)}{(\mathrm{d} n_{\rm as}/\mathrm{d} x)_{x=0}}.
\end{equation}

The models described by the KKE and the LBE lie at opposite poles. The latter has been called ``inverse Brownian motion'' \cite{Keilson+Storer1952,Andersen+Shuler1964}, and  Hoare \cite{Hoare1971AdvChemPhys} refined the terminology by referrring to the former \textit{regular} Brownian motion. A diffusing particle will be called  a ``B-particle'' or  an ``L-particle'' according as it obeys the KKE or LBE. Neither of these models is regarded as too remote to resemble a real physical system; the KKE is widely believed to provide a serviceable description of the thermal wanderings of a large particle (such as a colloid suspended in a liquid), and the LBE has found numerous applications in the transport of photons through a turbid medium or of a neutron through a moderator.

Ziff investigated the flux of particles to a single trap for two systems, only the first of which will be discussed here, namely that in which the diffusing particles execute jumps (named the ``Rayleigh flight'' by Ziff), all of the same length $l$, in three dimensional space, and are absorbed by a spherical trap of radius $R$; a particle undergoing this type of random walk will be called an R-particle. Smoluchowski carried out a detailed analysis of R-particles and wrote a paper of considerable pedagogical value\cite{Smoluchowski1906Cracovie}, but he did not think of using the model when he turned his attention, some ten years later, to the trapping of diffusing particles by absorbing surfaces. Since the test particle of a ``Rayleigh gas" is a B-particle, I will refer to the model used by Ziff as a Pearson-Rayleigh random walk (P\&R-RW, for short), and a particle undergoing such a walk as a P\&R-particle or a ``Pearson-Rayleigh random walker''.
The LBE and the P\&R-RW (but not the KKE) has each an intrinisic unit of length, denoted here as $\ell$ and $l$, respectively. As their unit of length, B\&T [whose symbol $\gamma$ is replaced here by $\zeta$] chose the quotient $\Lambda\equiv (kT/m)^{1/2}/\zeta$, where $\zeta$ is the friction coefficient. ``This `velocity persistence length', they remarked, ``plays a role similar to that of the mean free path in kinetic theory.'' Though common, the choice is not unique; for solving the three-dimensional version of the KKE, the Trondheim group found it more convenient to use a different unit, $\delta\equiv (2kT/m)^{1/2}/\zeta$. The diffusion coefficient of a B-particle, we note, is uniquely defined as $D=kT/m\zeta$, and that of an L-particle as $\frac{1}{3}\overline{v}\ell$, where $\overline{v}=(8kT/\pi m)^{1/2}$ is the average thermal speed.

\subsection{Results for the particle density}
\label{subsec:resdensity}

Since the expression for $n(x)$ will depend on the equation used for finding it, we will distinguish between the results pertaining to the DE, KKE, and LBE by adding a superscript and write, for example, $n^\mathrm{(DX)}(x)$, (for first-order results) and $n^\mathrm{(X)}(x)$ (for exact results), with $\mathrm{X}=$ K or L. 

The particle density $n^{\rm (X)}$  in the Milne problem can be expressed as: 
\begin{equation}
\begin{split}
n^{\rm (X)}(x)&=A^{\rm (X)}\left [x+x_{0}^{\rm (X)} +\Upsilon^{\rm (X)}(x)\right ]\\
&=n^{\rm (X)}_{\rm as}(x)+n^{\rm (X)}_{\rm tr}(x),
\end{split}
\end{equation}
where
\begin{equation}
\begin{split}
n^{\rm (X)}_{\rm as}(x)&=A^{\rm (X)}[x+x_0^{\rm (X)}],\\
 \mbox{and}\quad n^{\rm (X)}_{\rm tr}(x)&=A^{\rm (X)}\Upsilon^{\rm (X)}(x);
\end{split}
\end{equation}
the transient terms, $n^{\rm (L)}_{\rm tr}(x)={\cal O}(e^{-x/\ell})$ and $n^{\rm (B) }_{\rm tr}(x)={\cal O}(e^{-x/\Lambda})$, are always negative. The values (exact and first-order) of $A^{\rm (X)}$ and $A^{\rm (DX)}$, and $x_{0}^{\rm (X)}$  and $x_{0}^{\rm (DX)}$ are listed below:
\begin{subequations}
\begin{align}
\noalign{\noindent\mbox{\textbf{LBE}\quad ($D={\textstyle\frac{1}{3}} \overline{v}\ell$)}}
\mbox{1st order:}\hspace{2pt}    A^{\rm (DL)}	&=|j|/D,		&x_{0}^{\rm (DL)}	&=2D/\overline{v}\nonumber\\&&&={\textstyle\frac{2}{3}}\ell 	\label{eq:ResultsL1}\\
\mbox{Exact:}\hspace{2pt}    A^{\rm (L)}		&=|j|/D,	&x_{0}^{\rm (L)}	&=0.7104\,\ell\hspace{2em} 			\label{eq:ResultsLE}\\[6pt]
\noalign{\noindent\mbox{\textbf{KKE}\quad ($D=kT/m\zeta$)}}
\mbox{1st order:}\hspace{2pt}     A^{\rm (DB)}		&=|j|/D,	&x_{0}^{\rm (DB)}   		&=2D/\overline{v}\nonumber\\&&&=1.25\Lambda	\label{eq:ResultsK1}\\
\mbox{Exact:}\hspace{2pt}    A^{\rm (B) }		&=|j|/D,			&x_{0}^{\rm (B) }		&=1.46\Lambda 			\label{eq:ResultsKE}
\end{align}
\end{subequations}

\section{Comments elicited by the solution to Milne's problem for B-particles}
When the Brownian analog of Milne's problem was first solved (numerically) by B\&T, they compared their value of $x_{0}^{\rm (B) }$ with the known result for $x_{0}^{\rm (L)}$, and remarked:\par
\smallskip
\leftskip=20truept
\setstretch{0.9}
{\small\noindent
Far from the wall the density increases linearly with distance, as one expects from the diffusion equation.  When this asymptotic solution is extrapolated across the boundary region it reaches zero not at the wall (as the solution of the diffusion equation with absorbing boundary would) but at some distance beyond it. The value we find for this ``Milne extrapolation length'' is, in the appropriate dimensionless units, approximately twice the value found in the radiative transfer problem. The density in the actual solution is everywhere lower than that of the extrapolated asymptotic solution, but of course it stays finite at the wall.}\par
\medskip
\leftskip=0pt
\setstretch{1.04}
\leftskip=0truept
\smallskip 
A reader of this passage will be left with the impression that $n^{\rm (B) }_{\rm as}(x)$ disagrees, not only with the density calculated by using the DE (together with a BC that sets the particle concentration at the wall equal to zero), but also with $n^{\rm (L)}_{\rm as}(x)$.  The poor performance of \textit{this} DE-solution is easily understood, and was \textit{immediately} recognized by Burger [but no one else, until much later] in a paper of great of power and packed with physical insight \cite{KRN2024Colossal}: the boundary conditon $n(0)=0$ (used by Smoluchowski) cannot be strictly valid, since $|j|=n\overline{v}$ is finite and $\overline{v}$ can never become infinite.  But the discrepancy between $x_{0}^{\rm (L)}$ and $x_{0}^{\rm (B) }$ is, if genuine, rather perplexing; whether it is real or not can be ascertained only after one has found some means of relating the two units of length, $\ell$ and $\Lambda$, there being no grounds for equating the two. 

\subsection{Length scales for Brownian moton and its inverse}
As we are portraying the same physical system by two different models, it stands to reason that the two length scales, $\Lambda$ and $\ell$, must be calibrated against an expression (with the dimension of length) involving parameters that are common to both models; for the system at hand, these parameters are $D$ and $\overline{v}$.
On setting $A^{\rm (X)}=1=A^{\rm (DX)}$, we get ${1\over 3}\ell\ \overline{v}=D=(kT/m\zeta)$, from which ensues the desired relation between $\Lambda$ and $\ell$:
\begin{align}\label{eq:fictitiousmfp}
\ell &= \frac{3D}{\overline{v}}
=\frac{3}{\overline{v}}\frac{kT}{m\zeta} \nonumber\\
&=\frac{3}{\overline{v}}\sqrt{\frac{kT}{m}}\Lambda=
\sqrt{\frac{9\pi}{8}}\Lambda \approx1.88\Lambda .
\end{align}
Converting the value of $x_{0}^{\rm (L)}$ given in Eq.~\ref{eq:ResultsLE}, we find that $x_{0}^{\rm (L)}=1.335/\Lambda$, quite close to $x_{0}^{\rm (B) }/\Lambda$.  

\end{multicols}

\vspace{-30pt}
\setlength{\unitlength}{1mm}
\begin{picture}(100, 10)(0,0)
\put(-7, 0){\line(1,0){183}}
\put(-7, 0){\line(0,1){2}}
\put(176, 0){\line(0,1){2}}
\color{black}
\end{picture}

\begin{figure}[!h]
\hspace{6em}
\begin{minipage}[h]{0.65\linewidth}
\includegraphics[width=\textwidth]{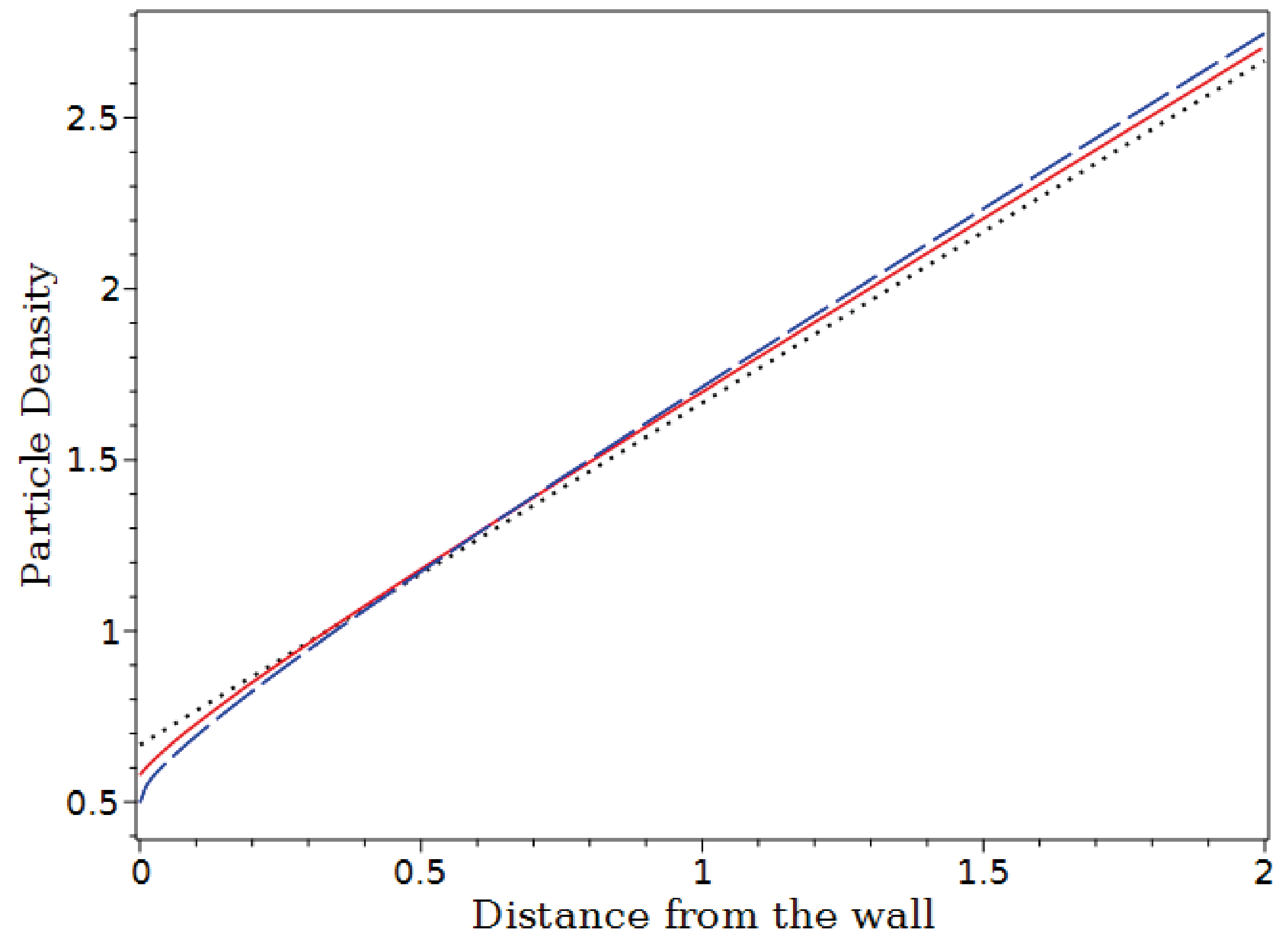}
\caption{Profiles of the particle density in the Milne problem (solid line) and its Brownian analog (dashed line); the distance from the wall is measured in units of the mean free path $\ell$. The dotted straight line is the result obtained by using the DE and the boundary condition $n(-x_0)=0$, where $x_0=2\ell/3$.}
\label{fig1}
\end{minipage}
\end{figure}

\vspace{-30pt}
\setlength{\unitlength}{1mm}
\begin{picture}(100, 10)(0,0)
\put(-7, 0){\line(1,0){183}}
\put(-7, 0){\line(0,1){2}}
\put(176, 0){\line(0,1){2}}
\color{black}
\end{picture}

\begin{multicols}{2}

\subsection{Profiles of particle density}

As a further check on the reliability of the conversion factor, we will look at the two density profiles plotted on the same scale. Figure~\ref{fig1} shows plots of near-exact approximations to $n^{\rm (B) }$  and $n^{\rm (L)}$, and compares these with the DE result $n^{\rm (D)}$ ($=n^{\rm (DL)}=n^{\rm (DB)}$); the sources of the data used for making the plots for $n^{\rm (B) }$  and $n^{\rm (L)}$ are described in Appendix \ref{secA1}.

Close to the absorbing boundary, the two densities ($n^{\rm (B) }$  and $n^{\rm (L)}$) differ appreciably, although not as much as one might anticipate before making such a comparison. At the coarse, coordinate-space level of the DE, the two transport equations, KKE and LBE, provide identical results. To distinguish between these diametrically opposed systems, one must go well beyond the two-term approximations to the distribution function used for reducing each of these equations to the DE.

\section{Flux to a spherical trap}
\label{sec:Flux2Tro}

For absorption of R-particles by a spherical trap of radius $R$, Ziff found the ``Milne extrapolation length'' to be $\approx 0.29795219 l$ for $0<l\leq 2R$. Since this result contradicts (at least at first sight) a great deal of work on closely related problems, including some carried out by the present author, and no resolution was offered in Ziff's paper \cite{Ziff1991Flux2trap}, or in two of its sparkling sequels \cite{MajumdarZiff2006JStatPhys,ZiffMajumdarComtet2007}, a probe into the discrepancy is warranted.

I will confine attention, for the most part, to the case $l/R\ll1$, which means that we can begin by looking at a system with plane symmetry, which reduces algebraic clutter to a great extent; still more simplification will result if I consider only the steady state. It will be instructive to study the diffusion of R-particles and L-particles moving with the same constant speed $u_0$. When it becomes necessary to distinguish between other quanities, superscripts will be added. 

Let us write down the expressions for the partial currents $j{\hspace{-2pt}\smash[t]{\strut}}_{\pm}(x)$, which denote the number of particles crossing unit area of an imaginary surface (placed at abcissa $x$) in the $\pm x$-direction. 
\vspace{-1em}
\begin{subequations}
\begin{align}   
\noalign{\noindent\mbox{L-particles:}}
j{\hspace{-2pt}\smash[t]{\strut}}_{-}(x)&=\frac{u_0}{2\ell}\int_{0}^{1}\hspace{-6pt}\mathrm{d}\mu\, \mu \int_{0}^{\infty}\hspace{-6pt} \mathrm{d}s\,\exp(-s/\ell) n(x+s\mu),\label{eq:jayM-L}
\\
j{\hspace{-2pt}\smash[t]{\strut}}_{+}(x)&=-\frac{u_0}{2\ell}\int_{-1}^{0}\hspace{-6pt}\mathrm{d}\mu\, \mu \int_{0}^{\infty}\hspace{-6pt} \mathrm{d}s\, \exp(-s/\ell) n(x+s\mu).\label{eq:jayP-L}
\end{align}
\end{subequations} 
\vspace{-2em}
\begin{subequations}
\begin{align}   
\noalign{\noindent\mbox{R-particles:}}
j{\hspace{-2pt}\smash[t]{\strut}}_{-}(x)&=\frac{u_0}{2l}\int_{0}^{1}\hspace{-2pt}\mathrm{d}\mu\, \mu \int_{0}^{l}\hspace{-2pt} \mathrm{d}s\, n(x+s\mu),\label{eq:jayM-R}
\\
j{\hspace{-2pt}\smash[t]{\strut}}_{+}(x)&=-\frac{u_0}{2l}\int_{-1}^{0}\hspace{-2pt}\mathrm{d}\mu\, \mu \int_{0}^{l}\hspace{-2pt} \mathrm{d}s\, n(x+s\mu).\label{eq:jayP-R}
\end{align}
\end{subequations} 

Since our aim \textit{here} is to obtain Fick's law (and thereby the DE), we now make the assumption that the spatial variation of $n^{\rm (D)}$, the  DE-approximation to the particle concentration, can be adequately described by a \textit{two-term} Taylor expansion:
\vspace{-4pt}
\begin{equation}\label{eq:2termTaylor}
n^{\rm (D)}(x+s\mu) = n^{\rm (D)}(x) + s\mu \frac{\mathrm{d}n^{\rm (D)}}{\mathrm{d}x}.
\vspace{-4pt}
\end{equation}
For the Milne problem, $\mathrm{d}^2n^{\rm (D)}/\mathrm{d}x^2$ vanishes identically; for a time-dependent problem, inclusion of the next term, $\frac{1}{2}\partial^2n^{\rm (D)}(x,t)/\partial x^2$, on the right-hand side of the time-dependent counterpart of Eq.~(\ref{eq:2termTaylor}) would still lead to Fick's law, but not to a boundary condition in which the particle concentration at a point on the absorbing surface is proportional to the gradient of the concentration at the point.  

Inserting the two-term expansion on the right-hand side of Eq.~(\ref{eq:2termTaylor}) into Eqs.~(\ref{eq:jayM-L}) and (\ref{eq:jayP-L}), we get
\begin{equation}\label{eq:jayMP-L}
j{\hspace{-2pt}\smash[t]{\strut}}_{\pm}(x)=\frac{u_0}{4}\left[n(x)\mp \frac{2\ell}{3}\frac{\mathrm{d}n}{\mathrm{d}x} \right].
\end{equation}
Whence follows Fick's law for L-particles:
\begin{equation}
\begin{split}
j\equiv j{\hspace{-2pt}\smash[t]{\strut}}_{+}(x) - j{\hspace{-2pt}\smash[t]{\strut}}_{-}(x)
&=-D^{(\rm L)}\frac{\mathrm{d}n}{\mathrm{d}x},\\
\mbox{with}\quad D^{(\rm L)}&=\frac{u_0\ell}{3}.
\end{split}
\end{equation}

We get a different result for R-particles,
\begin{equation}\label{eq:jayMP-R}
j{\hspace{-2pt}\smash[t]{\strut}}_{\pm}(x)=\frac{u_0}{4}\left[n(x)\mp \frac{l}{3}\frac{\mathrm{d}n}{\mathrm{d}x} \right],
\end{equation}
but this too leads to Fick's law, albeit with a different expression for the diffusion coefficient $D^{(\rm R)}$:
\begin{equation}\label{eq:D4R}
D^{(\rm R)}=\frac{u_0 l}{6}.
\end{equation}
Equation (\ref{eq:D4R}) does not contradict Ziff, who found  $D^{(\rm R)}={l^2}/{6\tau}$, because we also have the relation $u_0=l/\tau=l\nu$, where $\tau$ is the time taken for traversing a path of length $l$. For an L-particle, where the jump lengths show an exponential distribution, we have the relation $\sigma^2\equiv\overline{s^2}=2(\overline{s})^2 = 2\ell^2$, which can be combined with the relation $u_0=\nu \ell$ to get $D^{(\rm L)}={u_0\ell}/{3}=\nu\sigma^2/6=\sigma^2/6\tau$. 

We are thus led to the seemingly insipid but sublimely conciliatory conclusion that the expressions
\begin{equation}\label{eq:jayMP-both}
j{\hspace{-2pt}\smash[t]{\strut}}_{\pm}(x)=\frac{u_0}{4}\left[n(x)\mp \frac{2D^{(\rm X)}}{u_0}\frac{\mathrm{d}n}{\mathrm{d}x} \right]
\end{equation}
are valid both for L-particles and for R-particles. The conclusion is obvious because it merely echoes the following truism: no pair of expressions (involving only two terms) for $j{\hspace{-2pt}\smash[t]{\strut}}_{\pm}(x)$ will lead to Fick's law if it cannot be put in the form
\begin{equation}\label{eq:jayMP-obv}
j{\hspace{-2pt}\smash[t]{\strut}}_{\pm}(x)={\textstyle \frac{1}{4}}\overline{v}n\mp  {\textstyle \frac{1}{2}}D(\mathrm{d}n/\mathrm{d}x).
\end{equation}

\subsection{The boundary condition at the wall} 

We will define a black wall (located at $x=0$) by the relation $j{\hspace{-2pt}\smash[t]{\strut}}_{+}(0)=0$. Equations (\ref{eq:jayMP-L}) and (\ref{eq:jayMP-R}) immediately yield the following boundary conditions:
\begin{subequations}
\begin{align} 
n(0)&=\frac{2\ell}{3}\left.\frac{\mathrm{d}n}{\mathrm{d}x} \right|_{x=0} \quad\mbox{for L-particles},\\
n(0)&=\frac{l}{3}\left.\frac{\mathrm{d}n}{\mathrm{d}x} \right|_{x=0} \quad\mbox{for R-particles}.
\end{align}              
\end{subequations}
It follows from Eq.~(\ref{eq:jayMP-both}) that the boundary conditions above can be incorporated into the single relation
\vspace{-2pt}
\begin{equation}\label{eq:BC-both}
n(0)=\frac{2D}{\overline{v}}\left.\frac{\mathrm{d}n}{\mathrm{d}x} \right|_{x=0} ,
\vspace{-2pt}
\end{equation}
in which the superscript on $D$ has now been dropped, and $u_0$ has been replaced by $\overline{v}$, in order to allow for the possibility that all the test particles are not moving with the same speed.

If the wall absorbs only a certain fraction (say $\alpha$) of the incident particles, one only has to impose the condition $j{\hspace{-2pt}\smash[t]{\strut}}_{+}(0)=\alpha j{\hspace{-2pt}\smash[t]{\strut}}_{-}(0)$, which yields what may be called the mother of all boundary conditions (MBC):
\vspace{-2pt}
\begin{equation}\label{eq:MBC}
n(0)=\left(\frac{2-\alpha}{\alpha}\right)\Delta\left.\frac{\mathrm{d}n}{\mathrm{d}x} \right|_{x=0},\;\mbox{with}\; \Delta\equiv\frac{2D}{\overline{v}}.
\end{equation}

\subsection{The boundary condition for a spherical sink}
Citing four articles by Collins and co-workers, Ziff makes two remarks, stating, first, that the ``constant $\gamma$ [the linear extrapolation length] must be determined by empirical arguments'', and next, that ``to lowest order they [meaning various prescriptions by Collins et al] generally give the same result $\gamma =(l/3)[1+{\cal O}(\varepsilon)]$, \ldots'', where $\varepsilon\equiv l/R$. I am unable to see an unequivocal statment to this effect in any of the four papers cited by Ziff (his refs.~15--18). In their Eqs.~(3) and (4), reproduced here in my notation, Frisch and Collins \cite{FrischCollins1952JCP} provide the clearest statement of their BC:

\begin{equation*}
\alpha c(R,t)=\rho \left(\frac{\partial c}{\partial r} \right)_{r=R} \tag{F\&C-3}\label{eq:FC3}\\
\end{equation*}
\begin{equation*}
\rho =\frac{\displaystyle{\int_{0}^{\infty}s^2 \varphi(s)\hspace{1pt}{\mathrm d}s}}{\displaystyle{\int_{0}^{\infty}\hspace{-4pt}s\, \varphi(s)\hspace{1pt}{\mathrm d}s}}=\frac{\overline{s^2}}{\overline{s}},\tag{F\&C-4}\label{eq:FC4}
\end{equation*}

Since the $\alpha$-dependence is wrong (see below) and we are interested in a black sphere, we will set $\alpha=1$ in Eq.~(\ref{eq:FC3}). By examining ref.~15 of Ziff, a 1949 article \cite{CollinsKimball1949IndEngChem}, one can convince oneself that Eq.~(\ref{eq:FC4}) should be changed to
\begin{equation}\label{eq:CorrectRho}
\rho=\frac{1}{3}\frac{\overline{s^2}}{\overline{s}},
\end{equation}
but a general statement about $\rho$ (which is to be identified with the linear extrapolation length for a black sphere) is still beyond reach because\par
\smallskip
\leftskip=20truept
\setstretch{0.9}
{\small\noindent
 the ``jumps'' above discussed can be taken essentially as the path of the molecule between successive collisions. The persistence of velocity upon collision, however, causes the jump density function $\varphi(s)$ to be no longer spherically uniform but to depend upon the direction of the jump immediately preceding the jump under consideration in the manner of a Markov process.  However, \ldots  this effect can be accounted for by multiplying ${\langle s^2\rangle}$ by a correction factor slightly greater than unity. For convenience in this discussion, it will be assumed that this correction factor has been already incorporated in ${\langle s^2\rangle}$ and in the other moments of  $\varphi(s)$.}\par
\medskip
\leftskip=0pt
\setstretch{1.04}
\leftskip=0truept
\smallskip 
For the two cases of interest to us, namely L-particles and R-particles, we get $\rho=2\ell/3$ and $\rho=l/3$, in agreement with the calculations based on the plane symmetric system.

It is worth adding here that the jump model presented in the above cited 1949 article \cite{CollinsKimball1949IndEngChem} can be shown to imply the following boundary condition for a grey sphere:
\begin{equation}\label{eq:TBC}
\frac{2D}{u_0}\left(\frac{\partial n}{\partial r} \right)_{r=R}=\left(\frac{\alpha}{2-\alpha}\right) n(R,t)\quad ,
\end{equation}
but the inference did not appear in print until 1982 \cite{KRN+SW+KJM1982JPC}.

\subsection{The case of a small sphere: the Achilles heel of the ``Pearson-Rayleigh random walker''}
So far, we have not come across  any conclusion (emerging from Ziff's model) that cannot be reconciled with its counterparts from inverse and regular Brownian motion. However, Ziff's calculations revealed the extrapolation length $\gamma$ to be independent of $l/R$ for $0< l/R \leq 2$, in sharp contrast to the results found in neutron transport studies, which have been summarized by Sahni \cite{Sahni1966Black} and Williams \cite{Williams1988LinearExtrapol}, both of whom have presented their own calculations as well. When the Brownian analog of this problem was investigated, the coagulation rate constant showed an umistakable dependence on the value of $\Lambda/R$  \cite{KumarMenon1985JCP,WidderTitulaer1989JStatPhysV55,KRN1988Ark4}; an important conclusion that emerged from these investigations is worth stressing yet again: when moment methods are used for solving the KKE, attempts to obtain better results by increasing the number of moments will prove to be futile, because, beyond a certain order, convergence is lost.

\section{Molecular motion in liquids}

One of the most important and keenly controverted issues dicussed by Hildebrand in his thoughtful and purposely provocative note is the question whether concepts borrowed from the kinetic theory of gases or classical hydrodynamics can be legitimately applied to describe solute diffusion in liquid \cite{Hildebrand1971Science}.  The bulk of his criticism is to be found in the last two paragraphs, excerpts from which appear below:\par
\medskip
\leftskip=20truept
\setstretch{0.9}
{\small\noindent
Attempts to calculate absolute values of diffusivity have been made by starting from Stokes' law for a particle settling under the pull of gravity, and extrapolating over the long path to a molecule participating with its neighbors in aimless thermal motions that are never as long as the molecular diameter. Individual molecules are not impelled by a vector force that can serve to measure a ``coefficient of friction.'' Any such coefficient is fictitious.\par
If molecules were hard spheres, instead of electron clouds with imbedded nuclei, and were sufficiently far apart to justify speaking of binary collisions with linear free paths between them, the probable distance they could be expected to wander from their initial positions could be computed by the formula for a ``random walk.'' But polyatomic molecules that move less than 10 percent of their diameter require a more sophisticated mathematical formulation. They are in a continual state of soft, slow collision, with constant exchange between translational and internal energy. The random walk in this case is a slow, tipsy reel, without sudden changes of direction. The mathematical
problem involved \ldots 
}\par
\medskip
\leftskip=0pt
\setstretch{1.04}
\leftskip=0truept
\smallskip 

If one is content with the macroscopic description provided by the DE, concepts such as the mean free path and friction coefficient do indeed recede into the background, but not until we have derived a satisfactory boundary condition. If one wants to ascertain the validity of the macroscopic equations, for example the boundary condition for a particular problem, one is obliged to take a microscopic look.  Whether one uses the LBE (which speaks of trajectories between separable collisions) or the KKE (which envisages a medium with a given friction coefficient) for delimiting the domain of validity of the macroscopic description is quite immaterial, so long as one can convince oneself that one's conclusion does not rest on some idiosyncracy of the model; the means would justify the end provided that the end can be reached by some other means as well.  

\section{Concluding remarks}

Smoluchowski's work on colloidal kinetics has had a profound, though not purely beneficial, influence on the kinetics of colloidal coagulation and bimolecular chemical reactions; it raised, to be sure, the awareness that the diffusion equation (DE), supplemented by appropriate initial and boundary conditions, can be used for modelling a large variety of reacting systems, but it also instilled (in the minds of most of his readers) an unshakable conviction in the self-evidentness of his boundary condition for a perfectly absorbing (or black) surface. When introducing this fateful boundary condition, he added a footnote, the text of which reads \cite{Smoluchowski1916PhysZeit}:  ``Since the `speed' of Brownian motion is infinitely large for infinitely small distances, the adsorbing property of the wall must cause a complete removal of the particles from an infinitely thin layer adjacent to it.'' It was this boundary condition which enabled Smoluchowski to derive the expression $\Phi =4\pi R D n_0 [1+R/(\pi Dt)^{1/2}]$ for the rate at which the diffusing particles will coalesce to the surface of a black sphere of radius $R$; the quantity of prime interest for him was the long-time, stationary value $\Phi_{\rm{st}}=4\pi R D n_0$.  He made the proposal---without owning it to be a conjecture---that, when one is dealing with an imperfectly absorbing sink, the remedy is to multiply this expression by $\alpha$ \cite{Smoluchowski1916PhysZeit,Smoluchowski1916ZPC,Smoluchowski1917KZ}.

The claim---motivated by the search for a better boundary condition---that the Lorentz model is a useful tool for investigating bimolecular reactions in solutions became credible only after the publication of Burschka and Titulaer's numerical solution of the one-dimensional KKE \cite{Burschka+Titulaer1981JSPb}. Immediately prior to that, one critic expressed the ``community opinion'' by stating in a referee report (on an article co-authored by me) that the LBE ``is absolutely useless in dealing with transport in liquids'', and insisted that this task is best handled by solving the KKE. Only then did the need arise for comparing the length scales of inverse and regular Brownian motion. The Trondheim group has shown that inverse Brownian motion, regular Brownian motion and the BGK-model are indistinguishable at the DE-level, provided that one uses the appropriate BC \cite{KRN1982Ark7}, namely that stated in Eq.~(\ref{eq:TBC}). A Procrustean random walk model that allows no distribution of path lengths seems (to me) unphysical, much like the lattice model that informed Smoluchowksi's thinking (about the boundary condition at an absorbing surface), and has misinformed generations of students as well as aficionados of chemical kinetics.

Infinitely heavy B-particles, infinitely light L-particles, infinitely inflexible (about the constancy of their pathlenghts) R-particles are all fictions, but some fictions are more fruitful than others, and some are outright useless. Whether the fiction of R-particles will bear fruit (in the setting of diffusion-mediated reactions) or serve as a mere distraction remains to be seen.

\begin{appendices}

\section{Milne's problem: calculating the density profiles}\label{secA1}
The purpose of this appendix is to enable a reader of this article to generate the data used for plotting the density profiles shown in Fig.~\ref{fig1}.

The density data for L-particles were generated with the aid of a variational calculation \cite{KRN1993JQSRT}, in which $\ell$ was used a the unit of length.
The corresponding data for B-particles were computed by improving the results obtained by the Trondheim group through a half-range treatment \cite{KRN+KJM+SWPhysRevA1989}, in which the $N$th order approximation for the particle density $n$ was expressed in the form

\begin{equation}
n(x)=A\left[ (x+x_0)-\sum_{i=1}^{N-1}x_i \exp(-\lambda_i x/\Lambda)\right],
\end{equation}
and values of $x_0$ ($1.459877\Lambda$), $x_i$ and $\lambda_i$ (for $i=$1--8) resulting from a ninth-order approximation ($N=8$) were reported, and the values of $n(x)$ close to the wall were compared with those found by Marshall and Watson (M\&W) on the basis of their exact analytical treatment \cite{Marshall+Watson1987}. The improvements consists of three minor changes: the value of $x_0$ has been replaced by $x_0=1.460354\Lambda$ (the first seven figures of the exact result), one more term has been added, and the values of $x_i$ and $\lambda_i$ for the last three terms ($i=$7--9) have been optimised in a least-squares fit to the numbers in column (A) of Table 1 of M\&W. The complete set of $\{x_i,\lambda_i\}$ values (of mostly-analytical-partly-empirical origin) is displayed in Table~\ref{tab1}, the upper part of which is identical with Table II of ref.~\cite{KRN+KJM+SWPhysRevA1989}. For plotting the density of B-particles in Fig.~\ref{fig1}, the length scale was changed from $\Lambda$ to $\ell$.

\end{appendices}

\end{multicols}
\renewcommand{\arraystretch}{1.2}
\begin{table}[h]
\caption{Data for calculating near-exact values of the density of B-particles }\label{tab1}%
\hspace{5em}
\begin{tabular}{@{}llllllll@{}}
\toprule[0.3ex]
\quad$i=$		&1 				&2					&3					&4					&5					&6		 		\\
\hline\hline
$\lambda_i$    &1.000000  	&1.414231  	&1.737899		&2.108418		&2.797857		&4.359013	\\
$x_i$    			&0.097682  	&0.044722  	&0.030035		&0.035151		&0.048693		&0.064251  \\
\hline
\toprule[0.2ex]
\quad$i=$		&7		 			&8					&9					&		& 		&							\\
\hline\hline
$\lambda_i$    &9.703809		&31.872384 	&237.5727		&  	&  	&							\\
$x_i$    			&0.086451		&0.0480307		&0.069142 		&  	&  	&							\\
\bottomrule
\end{tabular}
\end{table}

\end{document}